\documentclass[11pt, a4paper, floatfix, fleqn]{revtex4}
\usepackage{vmargin,color} 
\usepackage[french,english]{babel} 
\usepackage[utf8]{inputenc} 
\usepackage{placeins}
\usepackage{graphicx,float,setspace,amssymb,amsmath,bm} 
 
\newcommand{\bmat}[1]{\boldsymbol{#1}}
\newcommand{\ga}{\gamma} \newcommand{\be}{\beta} \newcommand{\de}{\delta} 
\newcommand{\la}{\lambda} \newcommand{\al}{\alpha} 
\newcommand{\gD}{\Delta}  
\newcommand{\ep}{\epsilon} 
  
 \newcommand{\tend}{\rightarrow}

\newcommand{\stl}{\vskip 0.04\textheight}

\newcommand{\equa}[1]{\begin{eqnarray} \label{#1}}
\newcommand{\auqe}{\end{eqnarray}}
\newcommand{\auqenn}{\nonumber\end{eqnarray}}
\newcommand{\bloc}[1]{\begin{block}{#1}\begin{center}}
\newcommand{\colb}{\end{center}\end{block}}

\newcommand{\tab}[1]{\begin{tabular}{#1}}
\newcommand{\bat}{\end{tabular} \\ }
\providecommand{\abs}[1]{\left\vert#1\right\vert}

\newcommand{\dd}{\hat{m}_i\hat{m}_j -3 (\hat{m}_i \hat{r}_{ij})(\hat{m}_j \hat{r}_{ij})}

\begin{document}
\selectlanguage {english}
\title 
{{\bf Blocking temperature of interacting magnetic nanoparticles with
uniaxial and cubic anisotropies from Monte Carlo simulations. 
}} 
\author 
{V.~Russier }
\affiliation
{ICMPE, UMR 7182 CNRS and Université UPE \\ 
  2-8 rue Henri Dunant 94320 Thiais, France.}
\date{\today} 
\thispagestyle{empty} 
\begin {abstract}
  The low temperature behavior of densely packed interacting spherical single domain 
nanoparticles (MNP) is investigated by Monte Carlo simulations in the framework of
an effective one spin model. The particles are distributed through a hard sphere
like distribution with periodic boundary conditions and interact through the 
dipole dipole interaction (DDI) with an anisotropy energy including both cubic 
and uniaxial symmetry components. 
The cubic component is shown to play a sizable role on the value of the blocking 
temperature $T_b$ only when the MNP easy axes are parallel to the cubic easy
direction ([111] direction for a negative cubic anisotropy constant). 
The nature of the collective low temperature state, either ferromagnetic or 
spin glass like, is found to depend on the ratio of the anisotropy to the dipolar 
energies characterizing partly the  disorder in the system. 
\end {abstract}
\maketitle 
%
\section {Introduction}
Magnetic nanoparticles still arouse a great interest both on the fundamental point of view 
in nanomagnetism~\cite{skomski_2003,majetich_2006,skomski_2008,bedanta_2009,bedanta_2013} 
and in the field of applications especially in 
nanomedecine~\cite{pankhurst_2003,pankhurst_2009,owen_2012}.
Under a critical radius $r_{sd}$ depending on their chemical composition, determined by 
the ratio of the energy necessary to sustain a domain wall to the magnetostatic energy,
magnetic nanoparticles (MNP) are single domain objects (see for instance~\cite{dormann_1997,skomski_2003}). 
Typical values for $r_{sd}$ are 15~nm for Fe, 35~nm for Co, 30~nm for 
$\ga$-Fe$_2$O$_3$~\cite{dormann_1997,bedanta_2009}.
Under this critical size, the simplification due to the single domain character
leads to the effective one spin (or macrospins) models
where the MNP are represented as uniformly magnetized particles. 
As a consequence, the anisotropy energy is to be understood as an effective one, 
since no direct reference to crystalline defects and to spin canting can be considered,
and includes contributions stemming from the MNP shape, crystalline anisotropy or surface effects.
Although the local stucture of the MNP is then frozen, such effective one spin models (EOS) have 
shown to give results in agreement with experiments~\cite{russier_2012}.
In most cases the MNP are 
coated by a non magnetic layer making them exchange uncoupled. As a result, the
theoretical description of single domain magnetic nanoparticles assemblies faces 
mainly two difficulties: the long range nature of the dipolar interactions and the
one body magneto-crystalline anisotropy energy (MAE). The former is at the origin of 
collective behavior~\cite{bedanta_2009,bedanta_2013} at low temperature  
while the later leads 
to irreversibility, namely the so-called blocking phenomenon and hysteresis~\cite{dormann_1997}. 
Both the inter-particle interactions and the individual anisotropy influence the magnetic 
properties of the assemblies with different kind of signature, the former being essentially dependent
on the underlying structure while the later characterizes only the individual MNP. 
The importance of the dipole dipole 
interactions (DDI), for a given value of the MNP moment, namely for a given material and mean value 
of the diameter, is controlled by the inter-particle distances, {\it via} the MNP concentration in the 
assembly. In addition, the nature of the state induced at low temperature by the collective behavior 
resulting from the DDI depends strongly on the underlying structure and dimensionality because 
of the anisotropic character of the dipolar interaction. It can be in principle ferromagnetic or
anti--ferromagnetic for sufficiently ordered or concentrated systems~\cite{wei_1992,weis_1993,weis_2005}, 
or spin--glass like (SSG)~\cite{ayton_1997,petracic_2010,bedanta_2013} for more disordered ones as has been 
evidenced experimentally~\cite{petracic_2006,parker_2007,wandersman_2008,nakamae_2014,mohapatra_2015}.
We emphasize that the degree of structural disorder is of crucial
importance on this point and that this latter can be partly induced and/or modified 
by the MAE according to both its magnitude and the corresponding easy axes distribution.
On an other hand, the magnitude and the symmetry of the anisotropy terms depend on the crystalline 
nature of the material, the actual size, shape and surface characteristics of the 
MNP~\cite{coey_2010,bedanta_2009}. It is worth mentioning that the influence of both the DDI 
and the MAE on the magnetic properties are strongly temperature dependent: indeed, the MAE induced 
irresversibility behavior is expected below the blocking temperature, while the influence of 
the MAE on the room temperature magnetization curve is quite important only at sufficiently large 
values of the applied field~\cite{margaris_2012,russier_2013}. 
Concerning the MAE an important issue is the interplay between cubic and uniaxial 
symmetries~\cite{Usov_1997,geshev_1999,margaris_2012,sabsabi_2013,russier_2013,usov_2012,correia_2014}.
This has been studied in detail essentially for assemblies of non interacting~\cite{usov_2012,correia_2014} 
or weakly interacting particles~\cite{margaris_2012,sabsabi_2013}.
In the strong DDI coupling, corresponding to concentrated dried
powders of MNP this topic has been addressed for the magnetization curve at room 
temperature~\cite{russier_2013}.
One of the motivations of Refs.~\cite{usov_2012,russier_2013} was to 
examine the possibility to get evidence of a cubic component in the MAE from an 
experimentally observable property, a goal motivated by the usual experimental 
determination of the anisotropy constant from the blocking temperature once the MNP size 
is known through a Stoner-Wohlfarth relaxation law. Following this procedure, one
assumes {\it a priori} a uniaxial symmetry for the MAE while a large fraction
of works on MNP refer to particles made of materials  with cubic symmetry, 
as for instance the cubic spinel ferrites. Then the uniaxial
symmetry term can result from either the shape or surface contribution~\cite{skomski_2003}
but nevertheless the bulk cubic contribution cannot be simply ignored. Quite interestingly,
Garanin and Kachkachi~\cite{garanin_2003} and Kachkachi and Bonnet~\cite{kachkachi_2006} 
have shown that the surface anisotropy, when modeled from the Néel surface anisotropy 
model~\cite{neel_1954}, can be represented by a cubic contribution to the total MAE 
leading to an effective one spin model which strengthens the usefulness of the combined 
symmetries MAE study. 
In Refs.~\cite{soler_2012,neumann_2013} Monte Carlo simulations of magnetic properties including
the blocking temperature of thin films composed of $\ga$-Fe$_2$O$_3$ NP embedded in polyaniline
have been performed with NP interacting through DDI and characterized by two MAE terms of
uniaxial symmetry, stemming from shape and crystalline anisotropy respectively. 

In the present work we focus on the effect of combined anisotropies
of uniaxial and cubic symmetries. Conversely to the preceding paper,
where the equilibrium magnetization was concerned, we consider the
blocking temperature as determined from the so-called ZFC/FC curves.
As in our previous work~\cite{russier_2013}, we deal with concentrated systems with
high dipolar coupling and we deal only with MNP assemblies fixed in position
where the underlying structure is determined from a hard sphere like distribution.
A particular attention is paid on the nature of the frozen orientational 
state reached at low temperature, and we show the influence of the MAE induced disorder
on this point.

\section {Model} 
We model the assembly of spherical single domain MNP in the framework
of macro spin model. We neglect the size polydispersity and so
consider an assembly of dipolar hard spheres uniformly polarized with 
magnetization $M_s$ and diameter $d$. The uniaxial and cubic components of the
magneto-crystalline anisotropy, described by the corresponding single site interactions 
on the moment orientations, $\hat{m}_i$ (see equation (\ref{ener_2})) are characterized 
on the first hand by the anisotropy constants $K_u$ and $K_c$ and on the other hand by the 
uniaxial $\{\hat{n}\}_i$ and the cubic $\{\hat{x}_{\al\;=\;1,3}\}_i$ easy axes 
respectively. The distance of closest approach between MNP, say $d_{eff}\;=\;d\;+\;\gD$,
may differ from $d$ due to a non magnetic coating layer of thickness $\gD/2$ surrounding the MNP.
We focus on an assembly of MNP in a frozen disordered state.
The structure of the assembly is modeled by a true hard sphere configuration,
say $\bmat{R}\;=\{\vec{r}_i\}$, obtained
by a Monte Carlo evolution of hard sphere particles starting from a
face centered cubic (fcc) lattice at a fixed density. The resulting distribution is controlled
through the radial distribution function $g_{hs}(r)$ and more precisely the contact
value, $g_{hs}(d_{eff})$, related the hard sphere fluid pressure~\cite{hansen_2006}.
The hard sphere system thus defined is totally characterized by the volume fraction
$\Phi$~=~$(\pi/6)\rho{}d_{eff}^3$ where $\rho$ is the number of particles per unit volume.
We also introduce the MNP volume fraction, $\Phi_p$~=~$\Phi(d/d_{eff})^3$.
The particles are then fixed in position.
Periodic boundary conditions are used throughout
the paper. 
The total energy of the system which includes the (DDI), the uniaxial and the cubic 
contributions to the anisotropy energy (MAE) and the Zeeman energy respectively
\equa{ener_1}
E = E_{dd} + E_u + E_c + E_z
\auqe
is given, in reduced form after introducing a reference inverse temperature,
$\be_0$~=~$1/(k_{B}T_0)$, by
\equa{ener_2}
\be_0 E &=& \ep_d 
            \sum_{i < j}  \frac{\dd}{r_{ij}^{*3}}                     
          - \ep_{u} \sum_i (\hat{n}_i\hat{m}_i)^2
        - \frac{\ep_{c}}{2} \sum_i \sum_{\alpha} (\hat{m}_i\hat{x}_{\alpha i})^4
        - h \sum_{i} \hat{m}_i \hat{h} ~.                               \nonumber \\
\auqe
with the coupling constants and the reduced external field,
\equa{coupl_cste}
            \ep_{u} &=& \be_0 K_u v(d) ~;~  \ep_{c} = \be_0 K_c v(d)   \nonumber \\
            \ep_d   &=& \frac{\be_0 \mu_0}{4\pi} (\pi/6) M_s^2 v(d) (d/d_{eff})^3 
            ~\equiv~ \ep_d^{(0)}(d/d_{eff})^3 ~~;~~
      h = \be_0\mu_0 M_s v(d) H_a ~\equiv~ {H_a}{H_{ref}}              \nonumber \\
\auqe
The long ranged dipolar interactions (DDI) are treated
in the framework of the Ewald summation technique~\cite{allen_1987,frenkel_2002} and
the total expression including the latter can be found for instance in Ref.~\cite{weis_1993}.
In equations (\ref{ener_2}, \ref{coupl_cste}) hatted letters represent unit vectors, 
$r_{ij}^* = r_{ij}/d_{eff}$ and $v(d) = (\pi/6)d^3$ is the MNP volume.
We also introduce the reduced temperature, $T^*\;=\;T/T_0$.
Since we have in mind to model systems where no texturation is expected, such as powders 
or random close packed samples, we consider the MNP as randomly oriented one to each other.
Accordingly the cubic contribution, {\it a priori} related to the 
crystallographic orientations of the MNP's considered as nano crytallites (NC), is 
characterized by randomly distributed set of axes $\{\hat{x}_{\al}\}_i$. On the 
other hand, as in Ref.~\cite{russier_2013} the uniaxial easy axes are also randomly 
distributed but are either uncorrelated with or fixed to a given crystallographic 
orientation of the NC. In this later case, we consider the situation of uniaxial and 
cubic contributions acting in a constructing way. With a negative cubic anisotropy 
constant used throughout this work, the usual case for cubic spinel ferrites, 
this means uniaxial easy axes in the $[111]$ direction of the NC. 
A similar situation is obtained with $k_c\;>\;0$ and  uniaxial easy axes 
parallel to one of the cubic easy axes, namely $\hat{n}_i\;=\;\hat{x}_i$
(equivalently $\hat{y}_i$ or $\hat{z}_i$).

The determination of the blocking temperature is performed from the so-called
FC/ZFC magnetization curves which we simulate from Monte Carlo runs including
at each temperature step
2.5\;10$^4$ Monte Carlo steps (MCS) the last 1.25\;10$^4$ of which are used to 
perform the thermal averages. The ZFC curves are initiated from a true demagnetized 
low temperature state, which is obtained from a long Monte Carlo run, up to 
2\;10$^5$ MCS performed using a parallel tempering scheme~\cite{hukushima_1996} 
to overcome the expected slowing down behavior due to the strong dipolar coupling 
regime as well as the deep MAE potential wells.
Each ZFC and FC magnetization curve is determined from the average on a set 
(up to 48) of independent MC paths performed 
using either one or several different structural configurations (hard sphere 
distributions) This is is necessary in order to get an average over disorder, 
see equation~(\ref{m_spont}), due to both the random character of the easy axes distribution 
and the underlying hard sphere like structure. This procedure is performed using a 
parallel code where the paths are run simultaneously and the corresponding averages 
are calculated as a final step.

On an other hand, since we deal with concentrated system with a high
dipolar coupling, we expect the dipolar particles to present a collective
state which may be of ferro magnetic or spin glass character. To discriminate
from these possibilities we examine, at zero external field,
the spontaneous magnetization and the nematic order parameter $\la$ defined as 
the largest eigenvalue of the second rank tensor~\cite{levesque_1994}
\equa{def_la}
    \bar{Q} = \frac{1}{N} \sum_i \frac{1}{2} ( 3 \hat{m}_i\hat{m}_i - \bar{I}) ~ ,
\auqe
where $\bar{I}$ is the unit tensor. This is done as usual by using the so-called
conductive external conditions which prevent the occurrence of demagnetizing 
effect~\cite{allen_1987,weis_1993}. Such external conditions remain to consider 
the simulated system as immersed in an external medium of infinite permeability. 
For the simulation of the spontaneous magnetization, we also use the
parallel tempering scheme to avoid the important slowing down of the relaxation 
in the system related to the occurrence of the frozen collective state. 
The spontaneous magnetization $<M>$ is determined from the average of thermal
statistical mean values, $<M(\bmat{R}_n)>_T$ over a set of different hard sphere 
configurations $\bmat{R}_n$~\cite{klapp_2001}, namely 
\equa{m_spont}
<M>~= \frac{1}{N_R}\sum_{n=1}^{N_R} <M(\bmat{R_n})>_T .
\auqe
\section {Results}

In this work we do not aim to simulate a specific experimental system; instead
we focus on the qualitative trend expected for typical magnetic MNP with
iron oxides as a guide for the choice of the physical parameters, 
because of their importance in the field and the representativity of their physical 
properties ($K_c$ or $M_s$), in the family of oxide spinel ferrites 
excepted CoFe$_2$O$_4$.
Since these materials present a cubic structure, we estimate the MAE uniaxial component 
from the shape anisotropy, assuming an aspect ratio of say $\de$~$\simeq$~1.10 to 1.20.
Starting from $\mu_0M_s$~=~0.5\;T and $K_c$~=~-13\;kJm$^{-3}$ (-4.7\;kJm$^{-3}$), 
the room temperature values of the cubic anisotropic constant for magnetite (maghemite),
and setting $T_0$ to the room temperature we get a bare dipolar coupling constant 
ranging from $\ep_d^{(0)}$~=~1 to 8.0, $\abs{\ep_c}$~$\simeq$~1.65 (0.60) to 8.10 (2.95) 
and $\ep_u$~$\sim$~0.5 to $\sim$~7.25 for MNP with diameter in between $d$~=~10\;nm to 20\;nm. 
This corresponds to the range of validity of the macro spin models since they are limited 
to MNP whose radius is on the one hand larger
than some threshold value, say $\sim$~2~nm~\cite{dormann_1997}, under which the majority of atoms
lie in the surface and on the other hand smaller than the single domain critical size $r_{sd}$
recalled in the introduction.
In the following the parameters retained for the simulations are 
taken within the above ranges of values. We limit the DDI coupling to
$\ep_d\;\leqslant\;2$ which, for instance, corresponds to include a coating layer 
$\gD/2\;=\;2\;nm$ on magnetite MNP of $d$~=~16~$nm$ in diameter.
Since we expect, from results obtained in the superparamagnetic regime~\cite{russier_2013}, 
that the cubic contribution leads to a sizable effect only when $\abs{\ep_c}\;>\;3\;\ep_u$ 
at least in the presence of DDI we limit our simulations to 
$\abs{\ep_c}\;=\;4\;\ep_u$.
All the simulations are performed for a dense system characterized by $\Phi$~=~0.493
($\phi_p\;\simeq\;0.250$ if we include a coating layer $\gD/2\;\simeq\;d/8.0$)
with $N_p$~=~1728 particles in the simulation box.
\stl
\begin{table}[h]
\tab{ l l l l l}
      \hline
      $\ep_d$ ~&~ $\ep_u$ ~&~ $\ep_c$ ~&~ $\hat{n}_u$ ~&~ T$_b$ \\
      \hline
       0.0 ~&~ 2.00 ~&~  0.00  ~&~       ~&~ 0.218  \\
       0.0 ~&~ 2.00 ~&~ -8.00  ~&~ rand  ~&~ 0.270  \\
       0.0 ~&~ 2.00 ~&~ -8.00  ~&~ [111] ~&~ 0.304  \\ 
       1.0 ~&~ 0.60 ~&~  0.00  ~&~       ~&~ 0.257  \\
       1.0 ~&~ 2.00 ~&~  0.00  ~&~       ~&~ 0.400  \\
       1.0 ~&~ 2.00 ~&~ -8.00  ~&~ rand  ~&~ 0.410  \\
       1.0 ~&~ 2.00 ~&~ -8.00  ~&~ [111] ~&~ 0.510  \\
       1.0 ~&~ 3.40 ~&~  0.00  ~&~       ~&~ 0.580  \\
       2.0 ~&~ 1.40 ~&~  0.00  ~&~       ~&~ 0.557  \\
       2.0 ~&~ 2.00 ~&~  0.00  ~&~       ~&~ 0.643  \\
       2.0 ~&~ 2.00 ~&~ -8.00  ~&~ rand  ~&~ 0.659  \\     
       2.0 ~&~ 2.00 ~&~ -8.00  ~&~ [111] ~&~ 0.735  \\
       2.0 ~&~ 3.40 ~&~  0.00  ~&~       ~&~ 0.693  \\
      \hline    
\bat
\caption{\label{simul_tb}
 Blocking temperature in terms of the coupling constants of the model.}
\end{table}

Let us recall that the cubic axes ($\{\hat{x}_{\al}\}_i$) 
are randomly distributed and we limit ourselves to two different situations: the 
uniaxial easy axes are either uncorrelated to the cubic ones or are fixed in the 
referential of the later in such a way to maximize the global anisotropy effect,
namely coincide with the [111] direction of the local frame, 
$\hat{n}_i\;=\;(\sum_\al \hat{x}_{\al{i}})/\sqrt{3}$ for $\ep_c\;<\;0$. 
This will be referred to in the following as random or [111] easy axes distributions respectively,
implicitly referring to the local referential of the MNP.
According to our preceding results, we expect comparable results 
for $k_c\;<\;0$ and $\hat{n}_i$~=~[111]$_i$ and for $k_c\;>\;0$ with
$\hat{n}_i$~=~$\hat{x}_i$, considered in Ref.~\cite{correia_2014} for non interacting MNP.
 
\begin {figure}
  \includegraphics [width = 0.9\textwidth,  angle = -00 ]{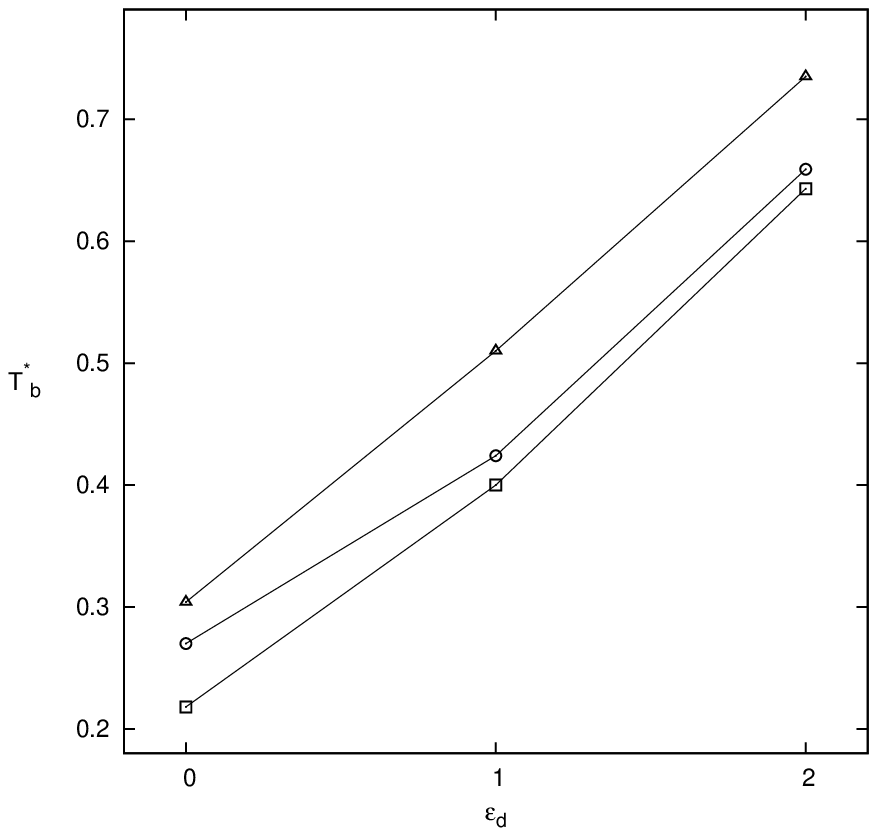}
  \caption {
  \label{tb_eu_2_ed0-2}
  Blocking temperature dependence on the dipolar interaction strength $\ep_d$,
   for $\ep_u$~=~2; $\ep_c$~=~0 (squares);
   $\ep_c$~=~-8 and $\hat{n}$ uncorrelated (circles); 
   $\ep_c$~=~-8 and $\hat{n}$ along the local [111] direction (triangles).
  }
  \end{figure}

\begin {figure}
  \includegraphics [width = 0.9\textwidth,  angle = -00 ]{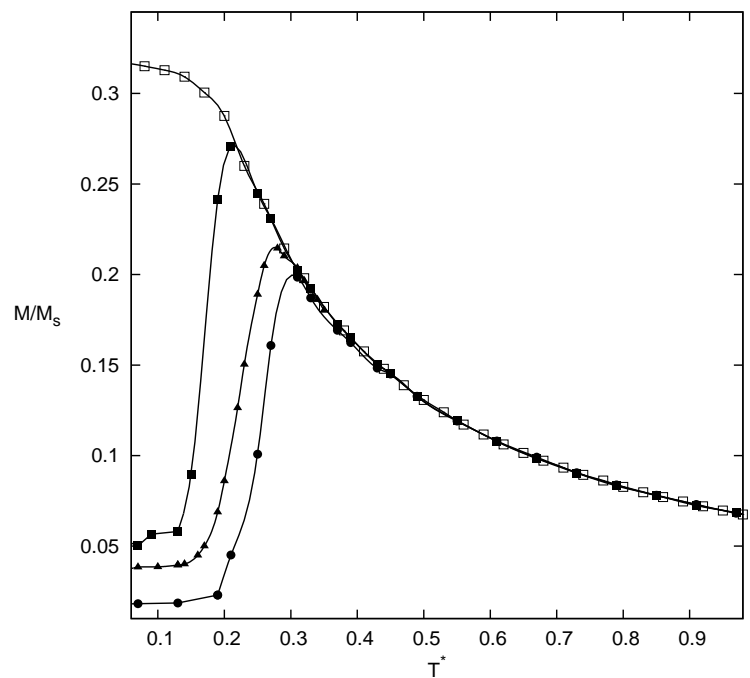}
  \caption {\label{hs_493_noddi_2}
  FC/ZFC magnetization curves for non interacting particles with $\ep_u$~=~2.
  Open squares: FC with $\ep_c$~=~0. 
  Solid symbols : ZFC curves with $\ep_c$~=~0 (squares); -8 and $\hat{n}$ uncorrelated (triangles);
  $\hat{n}$~=~[111] (circles).
  The lines are guides for the eye.
  }
  \end{figure}

\begin {figure}
  \includegraphics [width = 0.9\textwidth,  angle = -00 ]{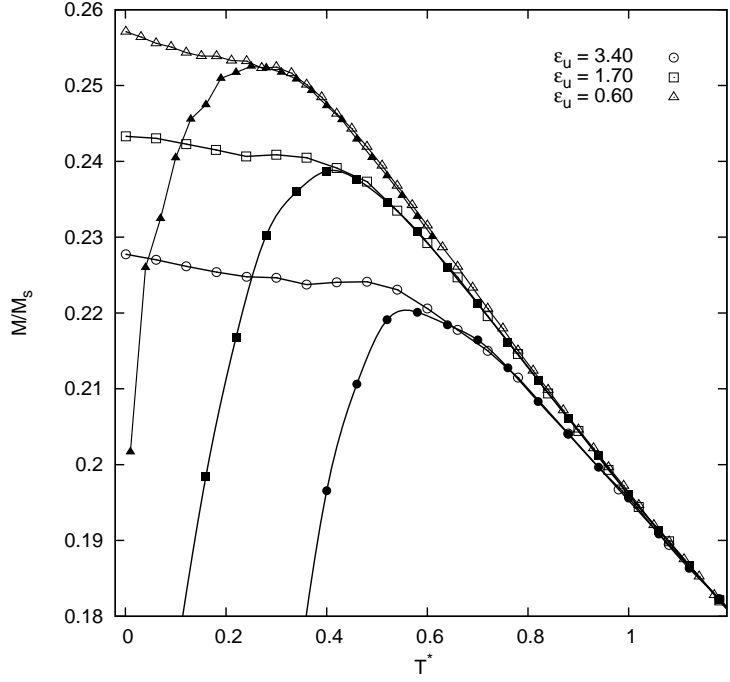}
  \caption {
    \label{hs_493_fc_zfc_ed1_ec0}
    FC/ZFC magnetization curve. $\ep_d$~=~1.0; $\ep_c$~=~0. Open symbols: FC curves; solid symbols: ZFC curves. The lines are guides to the eye.
  } \end{figure}

\begin {figure}
  \includegraphics [width = 0.9\textwidth,  angle = -00 ]{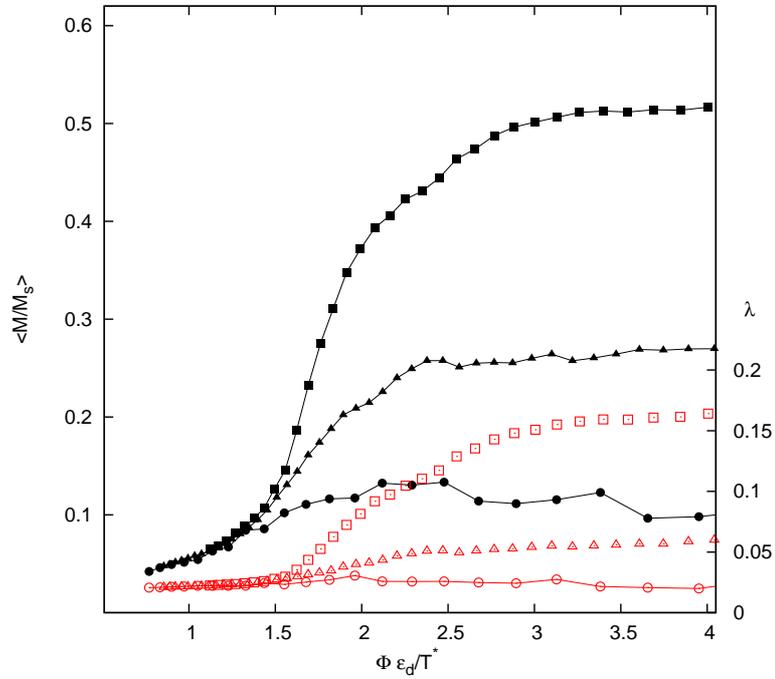}
  \caption {\label{pol_hs_493}
  Spontaneous polarization, left hand scale and solid symbols, and nematic order parameter $\la$, 
  right hand scale and open symbols, 
  in terms of $\ep_d/T^*$. Periodic boudary conditions with conducting external conditions.
  $\ep_c$~=~0.
  Squares: $\ep_u$~=~0; triangles: $\ep_d$~=~1.0 and $\ep_u$~=~0.60; 
  circles: $\ep_d$~=~2.0 and $\ep_u$~=~2.0.
  }
  \end{figure}

The results obtained for the blocking temperature are given in table \ref{simul_tb} and in 
figure~(\ref{tb_eu_2_ed0-2}).
First of all, we confirm the well known increase of $T_b$ with the DDI~\cite{garcia-otero_2000,vargas_2005}. 
The corresponding FC/ZFC curves are displayed in figures (\ref{hs_493_noddi_2}) 
and (\ref{hs_493_fc_zfc_ed1_ec0}) in the absence of cubic anisotropy for
non interacting and interacting MNP, with $\ep_d$~=~1.0, respectively.
The plateau like behavior of the FC magnetization curve at $T\;<\;T_b$, known to result from
the collective behavior induced by the DDI is clearly seen. 
At the qualitative level, the FC magnetization curves displayed in figure (\ref{hs_493_fc_zfc_ed1_ec0}) 
are quite similar, the main difference being the value taken by $T_b$; nevertheless we note that when
the MAE coupling constant takes a rather small value, $\ep_u$~=~0.60 and 1.40 for $\ep_d$~=~1 
and 2 respectively, {\it i.e.} $\ep_u/\ep_d\;\leqslant\;0.70$, the FC magnetization for $T\;<\;T_b$ 
still increases slightly when $T$ decreases, while for higher values corresponding to 
$\ep_u/\ep_d\;\geq\;1$, the FC curves present first a true plateau with possibly a small dip.
The latter behavior is known as the signature of a super spin glass 
state~\cite{de-toro_2013,petracic_2006,parker_2007,nakamae_2014}. Therefore, we are 
lead to conclude that when the ratio $\ep_u/\ep_d$ increases beyond a threshold value
$\ep_u/\ep_d$~=~$k_s$~$\sim$~0.6, a super spin glass state is reached at low temperature while, 
when $\ep_u/\ep_d\;<\;k_s$ the low temperature collective state may be closer to a ferromagnetic 
state expected for $\ep_u$~=~0 at least in the monodisperse case and for the high value of the 
volume fraction used here~\cite{klapp_2001,ayton_1997}. 
Since the easy axes are randomly distributed the ratio $\ep_u/\ep_d$ measures the degree
of disorder introduced in the system in addition to the structural one resulting from the
hard sphere like MNP distribution and thus is expected to bring the system from the ordered
ferromagnetic phase to the disordered super spin glass one.
Although the precise determination of the nature of the collective state at low temperature 
is beyond the scope of the present paper, we strengthen the above interpretation from the calculation 
of the spontaneous magnetization and of the nematic order parameter at zero field, given by equation 
(\ref{def_la}). The result is displayed in figure (\ref{pol_hs_493})
for $\ep_u/\ep_d$~=~0, 0.6 and 1.0 in terms of the relevant variable ($\Phi\ep_d/T^*$). 
The spontaneous magnetization curves versus $\Phi\ep_d/T^*$ in the cases $\ep_d$~=~1, 
$\ep_u$~=~0.6 and $\ep_d$~=~2, $\ep_u$~=~1.4 respectively are very similar. 
We see that in 
the absence of anisotropy a clear ferromagnetic phase is reached, while for $\ep_u/\ep_d$~=~0.6, 
although the spontaneous magnetization does not vanish, $\la$ takes a quite small value and no 
clear jump is obtained conversely to the $\ep_u$~=~0 case. It is worth recalling that 
pure dipolar hard spheres ($\ep_u\;=\;\ep_c\;=\;0$) located at the nodes of a well ordered 
{\it cfc} lattice, present a ferromagnetic phase characterized by both 
$<M/M_s>\;\tend\;1$ and $\la\;\tend\;1$ when $T^*\;\tend\;0$~\cite{weis_1993,weis_2005} 
and, more precisely, to compare with the results displayed in figure (\ref{pol_hs_493}) we 
get $<M/M_s>\;=\;0.91$ and $\la\;=\;0.75$ for ($\Phi\ep_d/T^*$)~=~4~\cite{russier_2015}. 
The spontaneously magnetized phase at zero field we get for $\ep_u$~=~0 is comparable to the 
disordered ferromagnetic phase obtained in~\cite{klapp_2001} on a quite similar system. 
We have checked from simulations with 864 or 2048 particles that in this later case, 
$<M>$ and $\la$ depend only very weakly on the system size, which shows that we are not dealing
with a large but finite cluster ferromagnetically correlated. 
Then either for $\ep_u/\ep_d\;>\;0.7$ with $\ep_c$~=~0 or in general $\ep_u\;\neq\;0$ 
and $\ep_c\;\neq\;0$ both the spontaneous magnetization and $\la$ vanish.  

\begin {figure}
  \hskip -0.25\textwidth
  \includegraphics [width = 0.76\textwidth,  angle = -00 ]{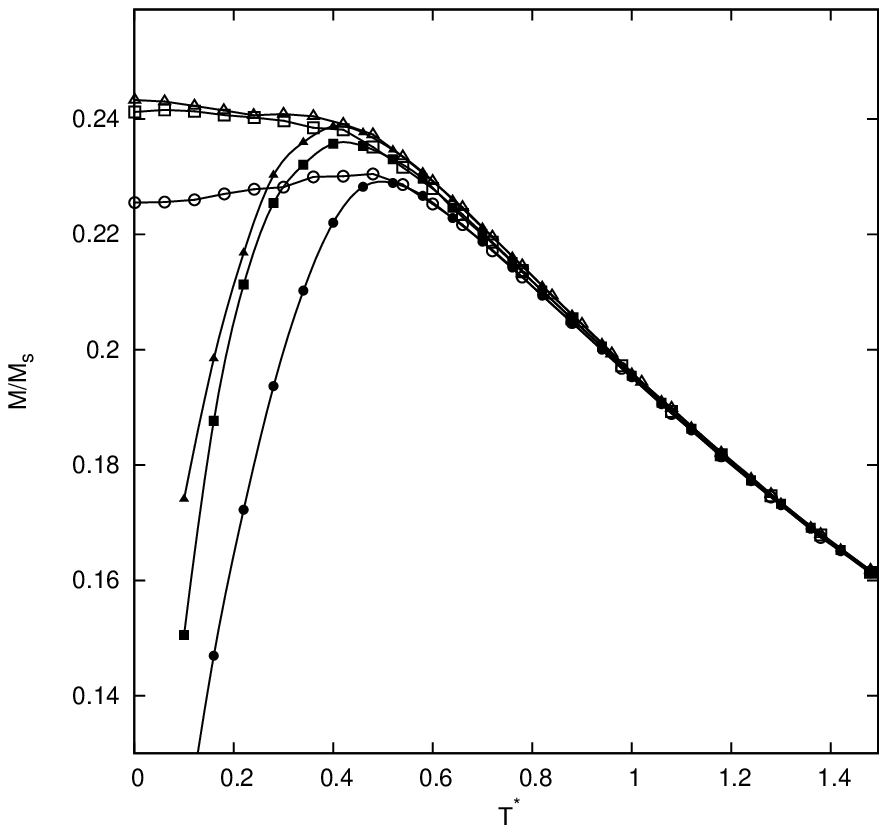}
  \hskip -0.28\textwidth
  \includegraphics [width = 0.76\textwidth,  angle = -00 ]{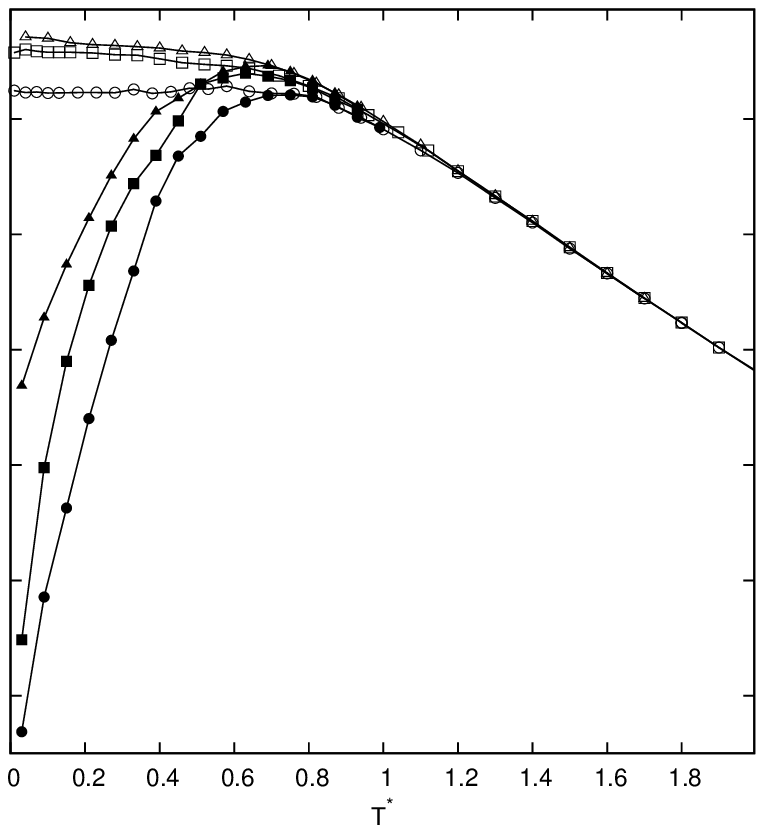}
  \caption {\label{hs_493_ed_eu_ec}
  FC (open symbols) and ZFC (solid symbols) magnetization curves.
  Left~: $\ep_d$~=~1.0 and $\ep_u$~=~1.70.
  Triangles: $\ep_c$~=~0; circles : $\ep_c$~=~-6.8 and $\hat{n}$~=~[111];
  squares: $\ep_c$~=~-6.8 and $\hat{n}$ uncorrelated. 
  Right~: $\ep_d$~=~2.0 and $\ep_u$~=~2.00.
  Triangles: $\ep_c$~=~0; circles : $\ep_c$~=~-8.0 and $\hat{n}$~=~[111];
  squares: $\ep_c$~=~-8.0 and $\hat{n}$ uncorrelated. 
  }
  \end{figure}

The effect of the MAE cubic contribution on $T_b$ depends strongly on the relative orientation
of the uniaxial and cubic easy axes. When the uniaxial easy axes are uncorrelated to the cubic ones,
the deviation of $T_b$ due to the cubic term is very small for MNP interacting through DDI and
is small but nevertheless significant in the non interacting case,
as can be seen in table~\ref{simul_tb} and on figure~\ref{tb_eu_2_ed0-2} and~\ref{hs_493_ed_eu_ec}. 
This is in agreement with what is found in the superparamagnetic 
regime for the behavior of the magnetization curve in the intermediate external field 
range~\cite{russier_2013}.
When the uniaxial easy axes $\{\hat{n}\}_i$ are set parallel to the [111] direction of the local 
MNP frames, the cubic contribution to the MAE has a sizable effect on $T_b$ both for non interacting 
and interacting MNP. 
The deviation $(T_b^*(\ep_c)~-~T_b(0))$ is then found nearly independent of $\ep_d$
which corresponds to a relative weakening of the influence of $\ep_c$ when 
$\ep_d$ is increased.
Indeed, still for $\{\hat{n}\}_i$ set parallel to the MNP [111] direction, 
if we consider $T_b$ in terms of the ratio $\abs{\ep_c}/\ep_u$ as is done in the non 
interacting case by Correia {\it et al.}~\cite{correia_2014}, we get from our simulations
$T_b(\abs{\ep_c}/\ep_u~=~4)/T_b(0)$~=~1.39, 1.26 and 1.14 for $\ep_d$~=~0, 1 and 2 
respectively. Notice that the result for the latter ratio in the non interacting case 
($\ep_d$~=~0) is very close to what Correia {\it et al.} have obtained~\cite{correia_2014} 
although the $T_b$ values are not strictly comparable since $T_b$ depend on the simulation 
'time' and moreover, Ref.~\cite{correia_2014} correspond to the easy axes in the [001] 
direction with $\ep_c\;>\;0$. 
The behavior of $T_b$ with respect to the cubic component in either the [111] or
the random uniaxial easy axes distribution, obtained in the present work
(see Table~\ref{simul_tb} and Fig.~\ref{hs_493_ed_eu_ec}),
is in agreement with that of the
hysteresis curves calculated by Usov and Barandiar\'{a}n~\cite{usov_2012}
in the sense that we show that both a strong cubic to uniaxial energies ratio
and conveniently correlated orientations of the two symmetries easy axes is necessary
for the cubic MAE contribution to significantly influence $T_b$.
The low temperature plateau of the FC magnetization curve is much more marked when the  
MAE cubic contribution does not vanish and especially for the [111] easy axes distribution.
Moreover, at low temperature ($\Phi\ep_d/T^*\;>\;4$), both the spontaneous 
magnetization and the nematic order parameter vanish when $\ep_c\;\neq\;0$.

Finally we compare the FC/ZFC and the resulting $T_b$ in the pure
uniaxial or cubic cases with comparable individual barrier heights.
This is done by noting that the uniaxial MAE barrier height is $\ep_u$
while the cubic MAE one is given by $-\ep_c/12$ with $\ep_c\;<\;0$ since
the moment must go through the saddle point in the [110] direction 
in order to jump from one [111] potential well to an other one.
Therefore we compare as an example the cases 
($\ep_u$~=~2,    $\ep_c$~=~0) with ($\ep_u$~=~0, $\ep_c$~=~-24) and 
($\ep_u$~=~1.40, $\ep_c$~=~0) with ($\ep_u$~=~0, $\ep_c$~=~-16.8) 
for $\ep_d$~=~0 and 2 respectively where we get $T_b$~=~0.22  and 0.29  on
the first hand and $T_b$~=~0.56  and 0.58  respectively. 
Only in the non interacting case the corresponding pure cubic $T_b$ values is 
significantly larger than the pure uniaxial one, while a quasi coincidence is obtained
for $\ep_d~\neq$~0. This is in agreement with the relative weakening due to the DDI of 
the $\ep_c$ contribution to $T_b$.

To conclude, in this work first of all we confirm at low temperature the results 
we obtained on large spherical clusters in the superparamagnetic regime, concerning 
the influence of the MAE cubic contribution. 
This contribution to the magnetization behavior is significant only when the MNP 
easy axis is locally fixed to the relevant MNP local orientation 
([111] or [100] for $\ep_c$~$<$~0 or $>$~0 respectively) 
and the corresponding deviation in $T_b$ is found nearly independent of $\ep_d$.
We also characterize at least at the qualitative level the nature of the frozen low 
temperature phase, which is found to present either a ferromagnetic or a spin-glass character 
depending on the value taken by the ratio $\ep_u/\ep_d$ contributing to the degree of disorder 
in the system for randomly distributed easy axes.
Conversely to our preceding work we consider only periodic boundary conditions 
and a true hard sphere distribution is used for the MNP structure with the limitation 
to the monodisperse case. 
%
\section {Acknowledgements}
We acknowledge very useful discussions with Dr. J.J.-~Weis from the LPT,
UMR~8627 of CNRS and Université Paris-Sud, Dr. I.~Lisiecki and Dr. J.~Richardi
from the MONARIS, UMR~8233 of CNRS and Université UPMC and Dr. S.~Nakamae
from the SPEC, UMR~2464 of CEA and CNRS.
This work was granted access to the HPC resources of CINES under the
allocation 2015-096180 made by GENCI.
We acknowledge financial support from the French research agency ANR 
grant ANR-CE08-007 for part of the  calculations performed in this work.
\FloatBarrier 

\FloatBarrier \eject
%
\end {document}